\documentclass[a4paper,9pt]{article}
\usepackage{pos}
\usepackage{bm}
\usepackage[export]{adjustbox}
\usepackage{graphicx}

\let\OLDthebibliography\thebibliography
\renewcommand\thebibliography[1]{
	\OLDthebibliography{#1}
	\setlength{\parskip}{1.4pt}
	\setlength{\itemsep}{0.6pt plus 0.5ex}
}

\title{Extraction of lattice QCD spectral densities from an ensemble of trained machines}

\author[1]{M.~Buzzicotti}
\author*[1]{A.~De Santis}
\emailAdd{alessandro.desantis@roma2.infn.it}
\author[1]{N.~Tantalo}
\affiliation[1]{Dept. of Physics and INFN, University of Rome Tor Vergata, Via della Ricerca Scientifica 1, I-00133 Roma, Italy}

\abstract{In this talk we discuss a novel method, that we have presented in Ref.~\cite{buzzicotti2023teaching}, to extract hadronic spectral densities from lattice correlators by using deep learning techniques. Hadronic spectral densities play a crucial role in the study of the phenomenology of strong-interacting particles and the problem of their extraction from Euclidean lattice correlators has already been approached in the literature by using machine learning techniques. 	A distinctive feature of our method is a model-independent training strategy that we implement by parametrizing the training sets over a functional space spanned by Chebyshev polynomials. The other distinctive feature is a reliable estimate of the systematic uncertainties that we obtain by introducing an ensemble of machines in order to study numerically the asymptotic limits of infinitely large networks and training sets. The method is validated on a very large set of random mock data and also in the case of lattice QCD data. }

\FullConference{The 40th International Symposium on Lattice Field Theory (Lattice 2023)\\
July 31st - August 4th, 2023\\
Fermi National Accelerator Laboratory\\}

\begin{document}
\maketitle

\section{Introduction}

The problem of the extraction of hadronic spectral densities from Euclidean correlators, computed from numerical lattice QCD simulations, has attracted a lot of attention over the years (see Refs.~\cite{Barata:1990rn,Jarrell:1996rrw,Nakahara:1999vy,Asakawa:2000tr,Burnier:2013nla,Hansen:2017mnd,Bulava:2019kbi,Hansen:2019idp,Kades:2019wtd,Bailas:2020qmv,Gambino:2020crt,Bruno:2020kyl,Horak:2021syv,Bulava:2021fre,Chen:2021giw,Wang:2021jou,Lechien:2022ieg,Bergamaschi:2023xzx,Rothkopf:2022fyo,Bulava:2023mjc,Karpie:2019eiq,Zhou:2021bvw,Boyda:2022nmh,Shi:2022yqw}). Indeed, from the knowledge of these key objects it is possible to extract all the information needed to study the scattering of hadrons and, more generally, their interactions. The problem of the extraction of spectral densities from lattice correlators is equivalent to that of an inverse Laplace-transform operation and, when dealing with a discrete and finite set of noisy input data, it is numerically ill-posed. Moreover, in the case of lattice field theory correlators further complications arise as lattice simulations have necessarily to be performed on finite volumes where the spectral densities are badly-behaving distributions. A regularization procedure is therefore mandatory and this can  be conveniently achieved by targeting the numerical calculation of \textit{smeared} spectral densities.

In Ref.~\cite{buzzicotti2023teaching} we have recently presented a new method for the extraction of smeared spectral densities from lattice correlators that is based on a supervised deep-learning approach. In order to develop an approach that can be used to obtain trustworthy theoretical predictions we have to address the following two pivotal questions
\begin{enumerate}
	\item is it possible to devise a \emph{model independent} training strategy?
	
	\item if such a strategy is found, is it then possible to quantify reliably, together with the statistical errors, also the unavoidable \emph{systematic uncertainties}?
\end{enumerate}

The importance of the first question can hardly be underestimated. Under the working assumption, supported by the so-called universal reconstruction theorems (see Refs.~\cite{HORNIK1989359,Goodfellow-et-al-2016,cybenko1989approximation}), that a sufficiently large neural network can perform any task, limiting either the size of the network or the information to which it is exposed during the training process means, in fact, limiting its ability to solve the problem in full generality. Addressing the second question makes the difference between providing a possibly efficient but qualitative solution to the problem and providing a scientific numerical tool to be used in order to derive theoretical predictions for phenomenological analyses. 

\section{Outline of the strategy and numerical setup}
\begin{figure}[h!]
	\begin{center}	
		\includegraphics[width=0.49\columnwidth,valign=c]{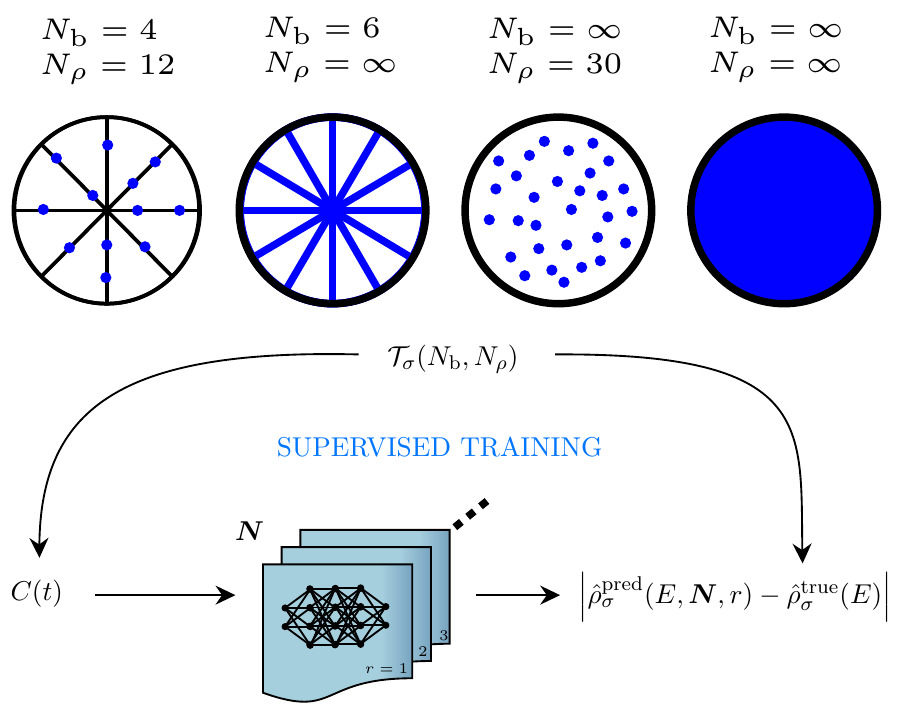}
		\includegraphics[width=0.5\columnwidth,valign=c]{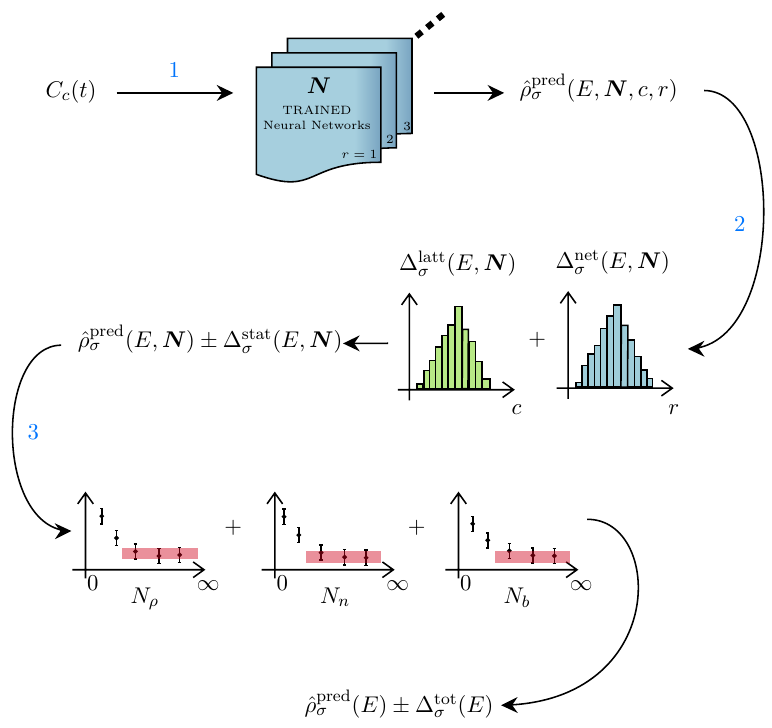}
		\caption{\small  
			\underline{\emph{Top-left}}:
			By introducing a discrete functional-basis, with elements $B_n(E)$, that is dense in the space of square-integrable functions $f(E)$ in the interval $[E_0,\infty)$ with $E_0>0$, any such function can exactly be represented as $f(E)=\sum_{n=0}^\infty c_n B_n(E)$. With an infinite number of basis functions ($N_\mathrm{b}=\infty$) and by randomly selecting an infinite number ($N_\rho=\infty$) of coefficient vectors $ \bm{c}=(c_0,\cdots, c_{N_\mathrm{b}})$, one can get any possible spectral density. This is the situation represented by the filled blue disk. If the number of basis functions $N_\mathrm{b}$ and the number of randomly extracted spectral densities $N_\rho$ are both finite one has a training set that is finite and that also depends on $N_\mathrm{b}$. This is the situation represented in the first disk on the left. The other two disks schematically represent the situations in which either $N_\mathrm{b}$ or $N_\rho$ is infinite.
			\underline{\emph{Bottom-left}}:
			We generate several training sets $\mathcal{T}_\sigma(N_\mathrm{b},N_\rho)$ built by considering randomly chosen spectral densities. These are obtained by choosing $N_\rho$ random coefficients vectors with $N_\mathrm{b}$ entries. For each spectral density  $\rho(E)$ we build the associated correlator $C(t)$ and smeared spectral density $\hat{\rho}_\sigma(E)$, where $\sigma$ is the smearing parameter. We then distort the correlator $C(t)$, by using the information provided by the statistical variance of the lattice correlator, and obtain the input-output pair $(C_\mathrm{noisy}(t),\hat{\rho}_\sigma(E))$ that we add to $\mathcal{T}_\sigma(N_\mathrm{b},N_\rho)$. We then implement different neural networks with $N_\mathrm{n}$ neurons and at fixed $\bm{N}=(N_\mathrm{n},N_\mathrm{b},N_{\rho})$ we introduce an \textit{ensemble of machines} with $N_\mathrm{r}$ replicas. Each machine $r=1,\cdots,N_\mathrm{r}$ belonging to the ensemble has the same architecture and, before the training, differs from the other replicas for the initialization parameters. All the replicas are then trained over $\mathcal{T}_\sigma(N_\mathrm{b},N_\rho)$ and, at the end of the training process, each replica will give a different answer depending upon $\bm{N}$.
			\underline{\emph{Right}}:
			Flowchart illustrating the procedure that we use to extract the final result. Here $C(t)$ represents the input lattice correlator that, coming from a Monte Carlo simulation, is affected by statistical noise. We call $C_c(t)$ the $c$-th bootstrap sample (or jackknife bin) of the lattice correlator with $c=1,\cdots, N_\mathrm{c}$. In the first step, $C_c(t)$ is fed to all the trained neural networks belonging to the ensemble at fixed ${N}$ and the corresponding  answers $\hat{\rho}_\sigma^\mathrm{pred}(E,\bm{N},c,r)$ are collected. In the second step, by combining in quadrature the widths of the distributions of the answers as a function of the index $c$ ($\Delta_\sigma^\mathrm{latt}(E,\bm{N})$) and of the index $r$ ($\Delta_\sigma^\mathrm{net}(E,\bm{N})$) we estimate the error $\Delta_\sigma^\mathrm{stat}(E,\bm{N})$) at fixed $\bm{N}$.  At the end of this step we are left with a collection of results $\hat{\rho}_\sigma^\mathrm{pred}(E,\bm{N})\pm \Delta_\sigma^\mathrm{stat}(E,\bm{N})$. In the third and last step,  the limits $\bm{N}\mapsto \infty$ are studied numerically and an unbiased estimate of $\hat{\rho}_\sigma^\mathrm{pred}(E)$ and of its error $\Delta_\sigma^\mathrm{tot}(E)$, with the latter also taking into account the unavoidable systematics associated with these limits, is finally obtained.
		}
		\label{fig:cartoon3}
	\end{center}
\end{figure}

The method that we have proposed in Ref.~\cite{buzzicotti2023teaching} (see FIG.~\ref{fig:cartoon3}) is built on two pillars
\begin{enumerate}
	\item the introduction of a \emph{functional-basis} to parametrize the correlators and the smeared spectral densities of the training sets in a model independent way;
	
	\item the introduction of the \emph{ensemble of machines} to estimate the systematic errors.
\end{enumerate}
In order to validate the method we consider both mock and real lattice QCD data and  choose the algorithmic parameters in order to be able to extract the hadronic spectral density from the lattice QCD  correlator $\bm{C}^\mathrm{latt}$, associated with the so-called \textit{R}-ratio, calculated on the B64 ensemble generated by the ETMC (see TABLE~I in Ref.~\cite{ExtendedTwistedMassCollaborationETMC:2022sta} and Ref.~\cite{Bernecker:2011gh,PhysRevD.107.074506}). We  then generate the needed mock data by starting from a model unsmeared spectral density $\rho(E)$ and by computing the associated correlator according to
\begin{flalign}
C(t)=\int_{E_0}^\infty \mathrm{d}\omega 
\frac{\omega^2}{12\pi^2}
\left[ e^{-t\omega}+e^{-(T-t)\omega}\right] \rho(\omega)\;,
\label{eq:inputC}
\end{flalign}
with $T=64a$ and $a=0.07957$ fm,  and the associated smeared spectral density according to
\begin{flalign}
\hat{\rho}_\sigma(E) = \int_{E_0}^{\infty} \mathrm{d} \omega \, K_\sigma(E,\omega) \rho(\omega)\;.
\label{eq:outputRho}	
\end{flalign}
We focus on  a Gaussian smearing kernel, $K_\sigma(E,\omega) = e^{-\frac{(E-\omega)^2}{2\sigma^2}}/\sqrt{2\pi \sigma^2} $, but  any other kernel can easily be implemented within the proposed strategy. Among the many computational paradigms available within the machine-learning framework, we opt for the most direct one and represent both the correlator and the smeared spectral density as finite dimensional vectors, that we use respectively as input and output of our neural networks. More precisely, the dimension of the input correlator vector is fixed to $N_T=64$, coinciding with the available number of Euclidean lattice times in the case of $\bm{C}^\mathrm{latt}$. The inputs of the neural networks are thus the 64-components vectors  $\bm{C}=\{C(a),C(2a),\cdots,C(64a)\}$. The output vectors are instead given by $\bm{\hat{\rho}_\sigma}=\{\hat{\rho}_\sigma(E_\mathrm{min}),\cdots,\hat{\rho}_\sigma(E_\mathrm{max})\}$. As in Ref.~\cite{ExtendedTwistedMassCollaborationETMC:2022sta}, we choose to measure energies in units of the muon mass, $m_\mu=0.10566$~GeV, and set $E_\mathrm{min}=m_\mu$ and $E_\mathrm{max}=24m_\mu$. The interval $[E_\mathrm{min},E_\mathrm{max}]$ is discretized in steps of size $m_\mu/2$. With these choices our output smeared spectral densities are the vectors $\bm{\hat{\rho}_\sigma}$ with $N_E=47$ elements corresponding to energies ranging from about $100$~MeV to $2.5$~GeV.

We treat $\sigma$, the width of the smearing Gaussian, as a fixed parameter by including in the corresponding training sets only spectral functions that are smeared with the chosen value of $\sigma$.  We consider two different values, $\sigma=0.44$~GeV and $\sigma=0.63$~GeV, that correspond respectively to the smallest and largest values used in Ref.~\cite{ExtendedTwistedMassCollaborationETMC:2022sta}.

\section{Model independent training set}

In the supervised deep-learning framework a neural network is trained over a training set which is representative of the problem that has to be solved. In our case the inputs to each neural network are the correlators $\bm{C}$ and the target outputs are the associated smeared spectral densities $\bm{\hat{\rho}_\sigma}$. As discussed in the introduction, our main goal is that of devising a model-independent training strategy. To this end, the challenge is that of building a training set which contains enough variability so that, once trained, the network is able to provide the answer, correct within the quoted  errors, for any possible input correlator.  In order to face this challenge we use the algorithmic strategy described in FIG.~\ref{fig:cartoon3}. In our strategy
\begin{itemize}
	
	\item the fact that the network cannot be infinitely large is parametrized by the fact that $N_\mathrm{n}$ (the number of neurons) is finite;   
	
	\item the fact that during the training a network cannot be exposed to any possible spectral density is parametrized by the fact that $N_\mathrm{b}$ (the number of basis functions) and $N_\rho$ (the number of spectral densities to which a network is exposed during the training) are finite;
	
	\item the fact that at fixed 
	\begin{flalign}\label{eq:N}
	\bm{N}=(N_\mathrm{n},N_\mathrm{b},N_{\rho})
	\end{flalign}
	the answer of a network cannot be exact, and therefore has to be associated with an error, is taken into account by introducing an ensemble of machines, with $N_\mathrm{r}$ replicas, and by estimating this error by studying the distribution of the different $N_\mathrm{r}$ answers in the $N_\mathrm{r}\mapsto \infty$ limit;
	
	\item once the network (and statistical) errors at fixed $\bm{N}$ are given, we can study numerically the $\bm{N}\mapsto \infty$ limits and also quantify, reliably, the additional systematic errors associated with these unavoidable extrapolations.
	
\end{itemize}
 For our numerical study we use the Chebyshev polynomials of the first kind as basis functions  and generate the unsmeared spectral densities that we use to build our training sets according to
 \begin{flalign}
 \rho(E;N_\mathrm{b})= \theta(E-E_0)\sum_{n=0}^{N_\mathrm{b}}c_n 
 \left[ T_n\left(x(E)\right)-T_n\left(x(E_0)\right) \right]\;,
 \label{eq:rho_cheby}
 \end{flalign}
 where $x(E)=1-2e^{-E}$ maps the energy domain $E\in [E_0,\infty)$ into $[-1,+1]$ in which the Chebyshev polynomials are naturally defined. The subtraction of the constant term $T_n(x(E_0))$ is introduced in order to include a threshold $E_0$ below which the spectral density is vanishing.  Each $\rho(E;N_\mathrm{b})$ that we use  to populate the training sets is obtained by choosing $E_0$ randomly in the interval $[0.2,1.3]$~GeV and by inserting in Eq.~(\ref{eq:rho_cheby}) the coefficients
 \begin{flalign}\label{p10}
 c_0 =r_0\;;
 \qquad
 \qquad
 c_n = \frac{r_n}{n^{1+\varepsilon}}\;,\quad n>0\;,
 \end{flalign}
 where the $r_n$'s are $N_\mathrm{b}$ uniformly distributed random numbers in the interval $[-1,1]$ and $\varepsilon$ is a non-negative parameter that we set to $10^{-7}$. Notice that with this choice of the coefficients $c_n$ the Chebyshev series of Eq.~(\ref{eq:rho_cheby}) is convergent in the $N_\mathrm{b}\mapsto \infty$ limit and that the resulting spectral densities can be negative.  Once $E_0$ and the coefficients $c_n$ that define $\rho(E;N_\mathrm{b})$ are given, the correlator $\bm C$ and the smeared spectral density $\bm{\hat \rho_\sigma}$ associated with $\rho(E;N_\mathrm{b})$ can be calculated by using  Eq.~(\ref{eq:inputC}) and Eq.~(\ref{eq:outputRho}). We generate $N_\rho$ pairs $(\bm C,\bm{\hat \rho_\sigma})$ and populate the training  sets at fixed $N_\mathrm{b}$ and $\sigma$, $\mathcal{T}_\sigma(N_\mathrm{b},N_\rho)$, with the pairs $(\bm{C}_\mathrm{noisy},\bm{\hat \rho_\sigma})$ where $\bm{C}_\mathrm{noisy}$ are random correlators extracted from the multivariate Gaussian distribution
 \begin{flalign}\label{eq:Multivariate}
 \mathbb{G}\left[\bm{C},\left(\frac{C(a)}{C_\mathrm{latt}(a)}\right)^2 \hat{\Sigma}_\mathrm{latt}\right],
 \end{flalign}
with center $\bm C$ and covariance $\hat{\Sigma}_\mathrm{latt}$, the covariance matrix  of the lattice correlator $\bm{C}^\mathrm{latt}$ mentioned above. 

In order to be able to perform a numerical study of the limits $\bm{N} \mapsto \infty$, we  generate with the procedure just described several training sets, corresponding to $N_\mathrm{b}=\{16,32,128,512\}$ and $N_\rho=\{50,100,200,400,800\}\times 10^3$. 

\section{Training setup and ensemble of machines} 

From the machine-learning point of view, we need to  implement  neural networks to solve a $\mathbb{R}^{N_T}\mapsto \mathbb{R}^{N_E}$ regression problem with $N_T=64$ and $N_E=47$.  For our analysis, we use feed-forward convolutional neural networks based on the LeNet architecture introduced in  Ref.~\cite{lecun1998gradient}. 
We study in details the dependence of the output of the neural networks upon their size $N_\mathrm{n}$ and, to this end, we implement three architectures that we call \textit{arcS}, \textit{arcM}  and \textit{arcL}. The architecture \textit{arcS} is described in full details in TABLE~\ref{tab:arcS}. \textit{arcM} and \textit{arcL} differ only for an increasing  number of maps in the convolutional layers for a total number of trainable parameters corresponding to 180871 and 371311. 

\begin{table}[t!]
	\begin{center}
	\includegraphics[width=0.5\textwidth]{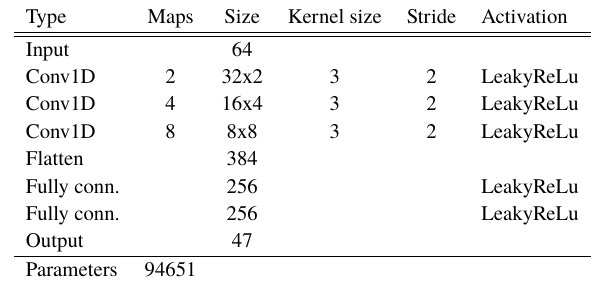}
	\end{center}
	\caption{\label{tab:arcS} \textit{arcS}: the smallest neural network architecture used in Ref.~\cite{buzzicotti2023teaching}. The architecture is of the type feed-forward and the structure can be read from top to bottom of the table. It consist of three 1D convolutional layers with an increasing number of maps followed by two fully connected layers.  The column denoted by 'Size' reports the shape of the signal produced by the corresponding layer. The stride of the filters is set to 2 in such a way that the dimension of the signal is halved at each 1D convolutional layer thus favouring the neural network to learn a more abstract, and possibly more effective, representation of the input data. The output layer has no  activation function in order not to limit the output range. 
	}
\end{table}
We train each machine with $N_\mathrm{n}$ neurons over the training set $\mathcal{T}_\sigma(N_\mathrm{b},N_\rho)$ 
by minimizing as loss function the Mean Absolute Error (MAE)
\begin{flalign}\label{eq:loss}
\ell(\bm{w}) = \frac{1}{N_\rho}
\sum_{i=1}^{N_\rho}\left\vert\bm{\hat{\rho}}_{\sigma}^{\mathrm{pred},i}(\bm{w})-\bm{\hat{\rho}}_{\sigma}^i\right\vert\;,
\end{flalign}
where  $\bm{\hat{\rho}}_\sigma^{\mathrm{pred},i}(\bm{w})$ is the output of the neural network in correspondence of the input correlator $\bm{C}_\mathrm{noisy}^i$.    We implement a Mini-Batch Gradient Descent algorithm, with Batch Size (BS) set to 32, by using the  Adam optimizer \cite{kingma2017adam} combined with a decaying learning rate. We adopt the standard early stop criterion to terminate the training by monitoring the validation loss (20\% of $N_\rho$).
As already stressed, at fixed $\bm N$ the answer of a neural network cannot be exact. In order to be able to study numerically the $\bm N\mapsto \infty$ limits, the error associated with the limited abilities of the networks at finite $\bm N$ \emph{has} to be quantified. To do this we introduce the \textit{ensemble of machines} by considering, at fixed $\bm N$, $N_\mathrm{r}=20$ machines with the same architecture. Each machine of the ensemble is trained by using a training set $\mathcal{T}_\sigma(N_\mathrm{b},N_\rho)$ obtained by starting from the same unsmeared spectral densities but with different noisy input correlators $\bm{C}_\mathrm{noisy}$.

\section{Validation and application}
\begin{figure}[t!]
	\begin{center}
	\includegraphics[width=0.5\columnwidth,valign=c]{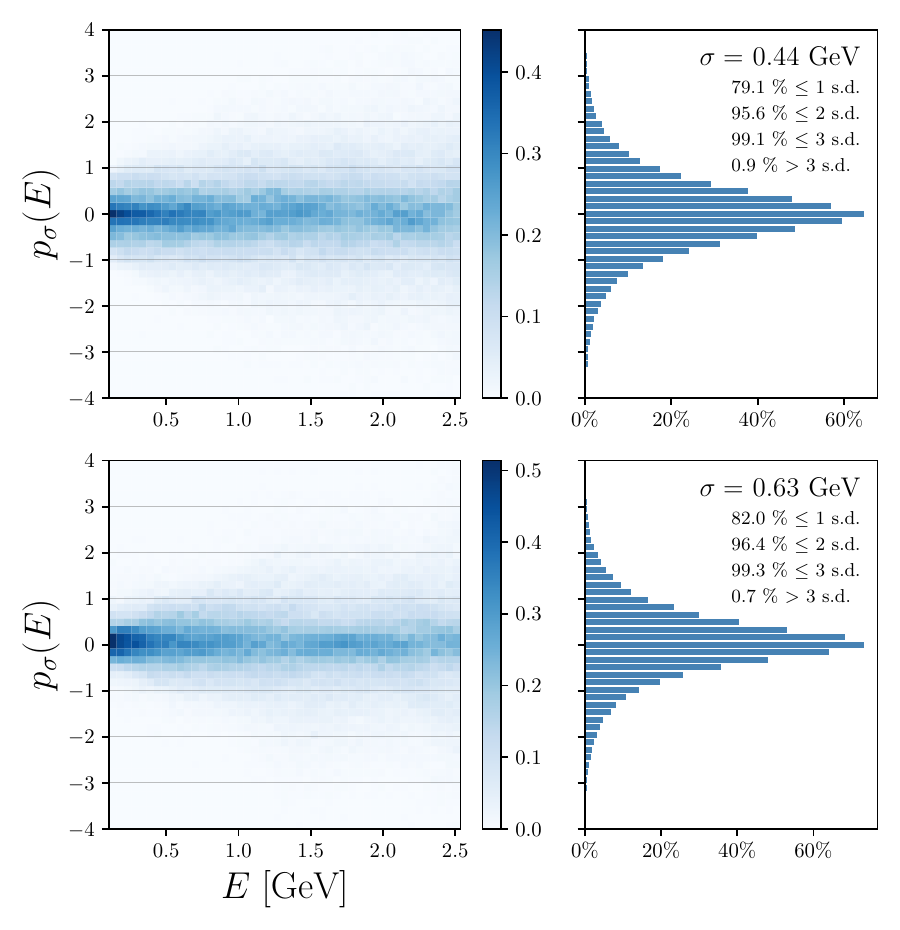}
	\includegraphics[width=0.49\columnwidth,valign=c]{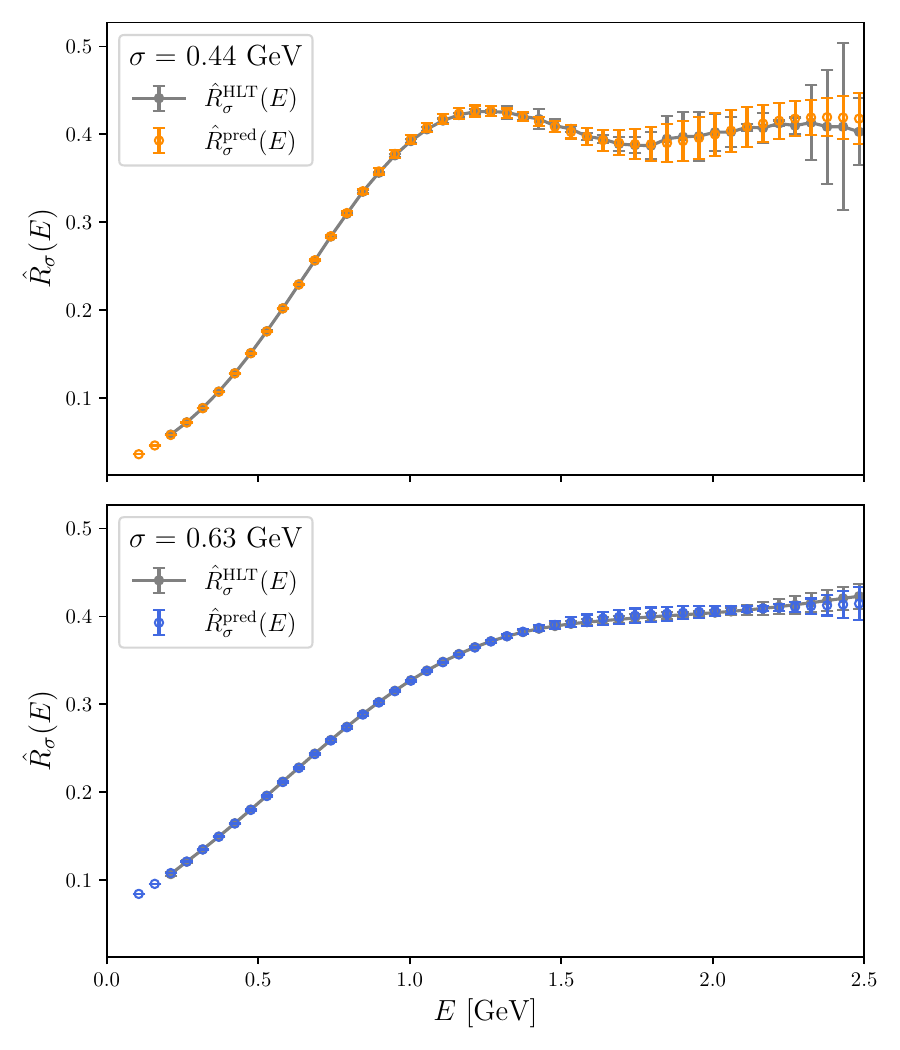}	\end{center}
	\caption{\small \underline{\emph{Left}}: Normalized density distribution of the significance defined in Eq.~(\ref{eq:signficance}) as a function of the energy. The distribution is calculated over 2000 validation samples generated according to Eq.~(\ref{eq:rho_peak_val}). Aside, the cumulative histogram of $p_\sigma(E)$ and related fractions of values that fall within 1, 2 or 3 standard deviations. \underline{\emph{Right}}: The results obtained with our ensembles of neural networks in the case of a real lattice QCD correlator are compared with those obtained with the HLT method (grey, Ref.~\cite{ExtendedTwistedMassCollaborationETMC:2022sta,Hansen:2019idp}) for the two values of $\sigma$. }\label{fig:peaks_sig}
\end{figure}
Once the ensemble of trained machines has been built, the procedure sketched in FIG.~\ref{fig:cartoon3} and illustrated in much more detail in Ref.~\cite{buzzicotti2023teaching}, is used to predict the smeared spectral density for any new correlator. In order to perform a stringent validation test of the proposed method we consider unsmeared spectral densities not belonging to the space spanned by the Chebyshev basis but generated according to
\begin{flalign}\label{eq:rho_peak_val}
\rho(E)=\sum_{n=1}^{N_\mathrm{peaks}} c_n \delta(E-E_n).
\end{flalign}
By plugging Eq.~(\ref{eq:rho_peak_val}) into Eq.~(\ref{eq:inputC}) and Eq.~(\ref{eq:outputRho}) we  calculate the correlator and smeared spectral density associated to each unsmeared $\rho(E)$. We  generate 2000 such unsmeared spectral densities by randomly generating $c_n$ and $E_n$. The trains of isolated Dirac-delta peaks are representative of unsmeared spectral densities that might arise in the study of finite volume lattice correlators and therefore represent a particularly challenging benchmarking set to validate the procedure. The left panel of FIG~\ref{fig:peaks_sig} shows the pull variable
\begin{flalign}\label{eq:signficance}
p_\sigma(E)=\frac{\hat{\rho}_\sigma^\mathrm{pred}(E)-\hat{\rho}_\sigma^\mathrm{true}(E)}{\Delta_\sigma^\mathrm{tot}(E)},
\end{flalign}
which is a quantitative estimate for the deviation of the predicted result from the true one within the total quoted error. $\Delta_\sigma^\mathrm{tot}(E)$ includes both statistical and systematic errors combined in quadrature. As it can be seen,  the final results are consistent with the true ones at the level of one standard deviation in about 80\% of the cases and below three standard deviations in more than 99\%. This is a reassuring evidence of the ability of our ensembles of neural networks to generalize very efficiently outside the training set and, more importantly, on the robustness of the procedure that we use to estimate the errors. \\

As application to real lattice QCD, we  consider a lattice correlator, measured by the ETMC, that has already been used in Ref.~\cite{ExtendedTwistedMassCollaborationETMC:2022sta} to extract the so-called strange-strange connected contribution to the smeared $R$-ratio by using the HLT method of Ref.~\cite{Hansen:2019idp}. The right panel of  FIG.~\ref{fig:peaks_sig} shows the comparison of $\hat{R}_\sigma^\mathrm{pred}(E)$ and $\hat{R}_\sigma^\mathrm{HLT}(E)$. The agreement between the two determinations is remarkably good over all the energy range. We stress that neither the new method proposed in Ref.~\cite{buzzicotti2023teaching} nor the HLT algorithm assume any prior information to predict the spectral density. The results presented in FIG.~\ref{fig:peaks_sig} are then unbiased determinations from first principles obtained from two completely different methods.

\section{Conclusions}

In Ref.~\cite{buzzicotti2023teaching} we  have proposed a new method to extract smeared hadronic spectral densities from lattice correlators. The method is based on supervised deep-learning and  the distinctive features are the implementation of a model-independent training strategy  and a reliable estimation of the systematics achieved by building an \textit{ensemble} of trained machines. 
We validate the method on a large set of mock data by measuring the failure rate of the ensemble of machines at the end of the procedure. In addition we  consider a true lattice QCD correlator obtained from a simulation performed by the ETM Collaboration. We extract the strange-strange connected contribution to the smeared $R$-ratio and compare the predictions obtained by using our ensembles of trained machines with the ones previously obtained by using the HLT method~\cite{ExtendedTwistedMassCollaborationETMC:2022sta,Hansen:2019idp}. The two determinations, obtained by using the two totally unrelated methods, show  a remarkably good agreement and a similar precision.

The proposed method  may be admittedly computationally demanding. However, on the basis of our experience, we are firmly convinced that a careful study of the different sources of systematic uncertainties is mandatory when dealing with machine-learning if the goal is   to compare theoretical predictions with experiments. In fact, the computational cost of the proposed method is the price that we have to pay for the reliability of the results. On the other end, our strategy, devised in the specific case of the extraction of spectral densities, can be fruitfully generalized to other applications and research fields.

\bibliographystyle{JHEP}
\footnotesize
\bibliography{bibli}

\end{document}